\documentclass[PRL,twocolumn,showpacs,preprintnumbers,amsmath,amssymb,superscriptaddress]{revtex4}
\usepackage{graphicx}
\usepackage{epstopdf}

  \def \In {$^{115}$In }
 \def\Ce {CeCoIn$_5 $ }
  \def\Tc {$T_{\rm c}$ }
   \def\Tf {$T^{\ast}$ }

\begin{document}

\title{Observation of Spin Susceptibility Enhancement  in the Possible FFLO State in CeCoIn$_5$}
\author{V. F. Mitrovi{\'c}}
\affiliation{Department of Physics, Brown University, Providence, RI 02912, U.S.A. }
\author{M. Horvati{\'c}}
\affiliation{Grenoble High Magnetic Field Laboratory, CNRS, B.P. 166, 38042 Grenoble
 Cedex 9, France}
\author{C. Berthier}
\affiliation{Grenoble High Magnetic Field Laboratory, CNRS, B.P. 166, 38042 Grenoble
 Cedex 9, France}
\affiliation{Laboratoire de Spectrom{\'e}trie Physique, Universit{\'e} J.  Fourier, B.P. 87, 38402 St. Martin d'H{\`e}res, France}
\author{G. Knebel}
\affiliation{D{\'e}partement de Recherche Fondamentale sur la Mati{\`e}re Condens{\'e}e, SPSMS, CEA Grenoble, 38054 Grenoble Cedex 9, France}
\author{G. Lapertot}
\affiliation{D{\'e}partement de Recherche Fondamentale sur la Mati{\`e}re Condens{\'e}e, SPSMS, CEA Grenoble, 38054 Grenoble Cedex 9, France}
\author{J. Flouquet}
\affiliation{D{\'e}partement de Recherche Fondamentale sur la Mati{\`e}re Condens{\'e}e, SPSMS, CEA Grenoble, 38054 Grenoble Cedex 9, France}

\date{Version \today}

\begin{abstract}
We report  \In nuclear magnetic resonance (NMR) measurements of   the heavy-fermion superconductor \Ce 
in the vicinity of the superconducting critical field $H_{c2}$ for a magnetic field applied perpendicular to the $\hat c$-axis. 
A possible inhomogeneous superconducting state, Fulde-Ferrel-Larkin-Ovchinnikov (FFLO),  is stabilized in this part of the phase diagram. 
In 11 T applied magnetic field, we observe   clear signatures of the two phase transitions:  
   the higher temperature one   to the homogeneous superconducting state and  the lower temperature phase transition    to a   FFLO state. 
   It is found that the spin susceptibility in the putative FFLO state is   significantly  enhanced  as  compared to  the  value in a homogeneous superconducting state.    
   Implications of this finding for the nature of the low temperature phase are discussed.
  \end{abstract}

\maketitle


 The recently discovered heavy-fermion superconductor \Ce with a quasi-2D electronic structure \cite{CedaDic}  offers a unique opportunity to investigate the interplay of unconventional superconductivity, magnetic fluctuations, quantum criticality, and non-Fermi liquid behavior. 
The experimental evidence indicates that \Ce is an unconventional superconductor with, most likely,  $d$-wave gap symmetry \cite{Izawa, Kohori, Curro01,Aoki04} below \Tc = 2.3 K.  
The $H_{c2}$ transition in this compound becomes 
 first-order below \mbox{$\simeq 0.7$ K}, for magnetic field $(H)$ applied both parallel and perpendicular to the $(\hat a \hat b)$ planes \cite{Izawa, Bianchi02,marphy02}, signaling that this  compound is  a superconductor subject to the   Pauli limit \cite{Maki66}. Thus,  it  offers for the first time the possibility to investigate the Fulde-Ferrell-Larkin-Ovchinnikov (FFLO) \cite{FFLOdis} state in an unconventional inorganic superconductor.
   This effect  is related to the fact that in a spin singlet superconductor  the destruction of superconductivity  by a magnetic field can be accomplished in two ways. Cooper pairs may break up either because of the coupling of the spin  degrees of freedom to the field (Pauli paramagnetism) or because of the effect of the field on the orbital  degree of freedom (vortices).
A novel FFLO phase is predicted to occur when Pauli pair-breaking dominates over the orbital effects  \cite{FFLOdis}.   In this new state the superconducting order parameter varies periodically in space and Cooper pairs acquire finite momentum, $|\vec q| \sim 2 \mu_B H/ \hbar v_F$.
 This concept, developed in the context of
solid state physics, is now highly topical for ultra-cold atomic
Fermi gases \cite{Zweierlein}.
  In addition, the study of the FFLO is of significant 
  interest   in elementary-particle physics  \cite{Casalbuoni}. 

 Single crystal samples of  \Ce appear to be ideal material for a microscopic investigation of the FFLO state owing to their high purity, layered structure, and  large spin susceptibility.  Many experiments \cite{Bianchi03, Radovan03, MakiPeak, Kakuyanagi05}  have identified a possible phase transition to a FFLO state. Even though  these measurements identified the phase transition within the Abrikosov superconducting (SC) state (henceforth referred to as uniform SC state) in the vicinity of the $H_{c2}$, the  nature of this fundamentally new SC state   remains to be elucidated.   

\begin{figure}[t]
\centerline{\includegraphics[scale=0.41]{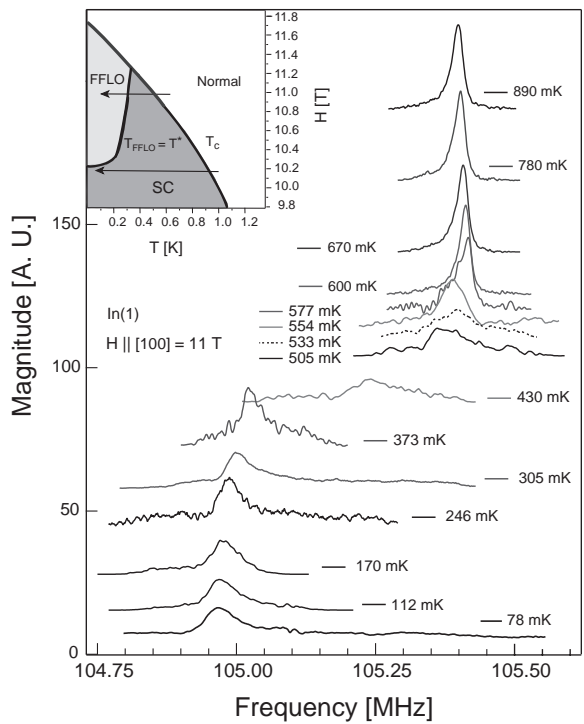}} 
 \vspace*{-0.2cm}
\caption[]{\label{Fig1}\small
NMR spectra of  $^{115}$In(1)  (central transition) as a function of temperature at 11 T 
magnetic field  applied  in the $[100]$ direction. 
The spectra are normalized to their corresponding areas and offset for clarity. 
The offset is proportional to the corresponding $T$.
Inset: Sketch of the $H - T$ phase diagram of \Ce for $H  \perp \hat c$  \cite{Bianchi03}. 
Arrows indicate  the fields in which spectra discussed in the paper were taken.
 }
 \vspace*{-0.3cm}
 \end{figure}

Here we report on a study of \In  magnetic shift in \Ce using nuclear magnetic resonance (NMR) in a magnetic field   applied parallel to $(\hat a \hat b)$   plane. 
The high field low temperature ($T$) part of the phase diagram was explored. 
The measurements probe  the local spin susceptibility.   At 11 T, they  clearly reveal two   phase transitions,  at \Tc and $T^{\ast}$    as indicated in the inset to 
\mbox{Fig. \ref{Fig1}}.  We show that  the     phase transition at  \Tc   is   associated with the transition to the uniform SC   state. 
We find that    below \Tf  the spin susceptibility is  enhanced  compared to  the  value in a uniform SC state, but reduced compared to the normal state.   
Within the phase we did not observe any normal state regions, interpreted 
as  evidence for the emergence of a spatially inhomogeneous SC state  \cite{Kakuyanagi05}. 
Thus, we 
argue that the state below \Tf  may be either a FFLO state that cannot simply be viewed as a stack of SC and normal regions,  or a state with a different spin fluctuation character, and consequently with possibly different order parameter   compared to the uniform  SC  state.

Single crystals of \Ce were grown by a flux method as described in \mbox{Ref. \cite{CedaDic}}. 
The crystal structure of \Ce is tetragonal and consists of alternating layers of CeIn$_3$ and 
CeIn$_2$. This leads to two inequivalent $^{115}$In sites per unit cell \cite{Moshopoulou, Curro01}. 
Our sample is believed to be 
of a good quality  since all observed In $(I=9/2)$ and Co $(I=7/2)$ lines are  narrow (in the normal state FWHM for Co line is $\simeq$ 12 kHz and $\simeq$ 20 kHz for In). 
Furthermore, the   quality of  the sample is confirmed by the observed  position  of the nine satellite lines for both In sites, which are in  agreement with the previously reported values \cite{Curro01}. 
 From the position of the satellites we infer that  no significant structural changes take place down to \mbox{70 mK} and that $H$ was aligned to better than 2$^\circ$ and 
 4$^\circ$ with respect to the sample's   $(\hat a \hat b)$ plane and  $\hat a$ axis, respectively.
 
%
%

The shift measurements of  the axially symmetric $^{115}$In(1) site   reported here were made on a large single crystal of CeCoIn$_5$. 
The   NMR spectra were recorded using a custom built NMR spectrometer and obtained, at each given value of $H$, from the sum of spin-echo Fourier transforms recorded at  each 10 (or 20) KHz interval. The shift was determined by the  
  diagonalization of the nuclear spin Hamiltonian  including the quadrupolar effects. 
  Temperature independent orbital  contribution ($\simeq 0.13 \%$) to the shift was subtracted.
     Below \Tc  the  demagnetization field lowers  the observed local field at the nucleus   site. Therefore,  an accurate     determination of the spin shift may be  precluded by this shielding field and the values of the extracted shift could be underestimated. The shift data presented in this paper  correspond to    a lower bound of the intrinsic  spin susceptibility  below $T_{\rm c}$.  
However, for   $H$ parallel to $(\hat a \hat b )$ planes the demagnetization factor is relatively small.

    The low $T$ environment  was provided by a $^3$He/$^4$He dilution refrigerator. 
   The RF coil was mounted into the mixing chamber of the refrigerator,  
   ensuring  good thermal contact. 
 To be able to span over a wide frequency range, we      used the `top-tuning' scheme in which 
  the  variable tuning delay line and matching capacitor
    were mounted outside the NMR probe. 
    For   $H$ values in the vicinity of  \mbox{11 T}  the central transitions of the two In sites are separated by only \mbox{$\simeq 400$ KHz}, which is of the order of the shift of the In(1) line at $T^{\ast}$. Therefore, it was necessary to check the satellite frequencies to assure that the correct site is followed as a function of $T$.  
     The sample was field-cooled. 
 In order  to avoid  heating of    the sample by  RF pulses we used an 
  RF excitation power much weaker than  usual and  repetition times of the order of 
   a tenth of a second to 
  several  seconds, depending on $T$.

 In  \mbox{Fig. \ref{Fig1}}  spectra of  $^{115}$In(1) as a function of  $T$ at \mbox{$H = 11$ T} are displayed. 
 Two phase transitions are evident. The first transition is from the normal state to the uniform SC  state at $T_{\rm c} \simeq 550$ mK and the second is from the uniform SC state to the supposed FFLO state between \mbox{$370 < T < 430$ mK}.  
 As \Tc is crossed a severe loss of the signal intensity   is observed. 
  The intensity   in the SC state at \mbox{$T \simeq 500$ mK} is reduced by an order of magnitude with respect to the normal state intensity. The loss of signal intensity is due to RF shielding by the superconducting currents. This observation confirms that  below \mbox{$\simeq 550$ mK} the sample is indeed in a  SC  state.
 
 We will discuss the possible microscopic nature of the low $T$   phase.
  The  lineshapes in the low $T$  phase  are clearly distinct from the ones at higher temperatures. 
  Nonetheless, these lineshapes are not consistent with a traditional vortex lattice field distribution  \cite{Brandt}.  That is,  the spectra  are  broaden well beyond the expectations for the vortex lattice lineshapes. 
  The progressive loss of the signal intensity (corrected for the Boltzmann factor), due to  the extra shielding of RF,   ceases on   lowering the $T$  below \mbox{370 mK}.
   However, the signal intensity   still remains significantly lower compared to the normal state intensity. 
   This would imply that below $T \simeq 370 \,{\rm mK}$ the SC order parameter, whatever its nature might be, is fully developed. 
   
 For $370 < T < 470$ mK the signal is extremely weak, almost indiscernible from the noise, as evident 
 in \mbox{Fig.  \ref{Fig1}}. 
 We   point out that the weakness of the signal is not only  due to RF penetration problems. The main cause of the loss of signal intensity is an enhancement of the spin-spin decoherence rate, $T_2^{-1}$. 
  A fast decoherence rate (shorter than the dead time  
  for the signal detection,  $\simeq 6 \rm{\mu s}$) 
  implies strong enhancement of a magnetic fluctuation component parallel  to  $H$ on approaching $T^{\ast}$.   
 It is well known that  enhanced fluctuations are precursory to magnetic transitions or  transitions associated with vortex dynamics. 
However,     the transition at \Tf cannot be associated with the changes in vortex dynamics,   since vortices must be in a solid state, {\it i.e.}  static on the NMR time scale, at all temperatures below $T_{\rm c}$, otherwise   the problem of RF penetration would not    be encountered throughout this $T$ range. 
Any static rearrangement of vortices at a fixed value of the applied field cannot account for the observed shift variation \cite{Brandt}.  
There is evidence of    abundant antiferromagnetic  spin fluctuations  in \Ce and of their coexistence  with uniform superconductivity \cite{Curro03, Sidorov02}. Therefore, it is conceivable to think that    magnetic fluctuations change  from antiferromagnetic to ferromagnetic-like or that the nature of antiferromagnetic fluctuations changes \cite{Takimoto02} at $T^{\ast}$. In the former scenario, this in turn could   cause the changes of the SC order parameter.

 %
   \begin{figure}[t]
\centerline{\includegraphics[scale=0.42]{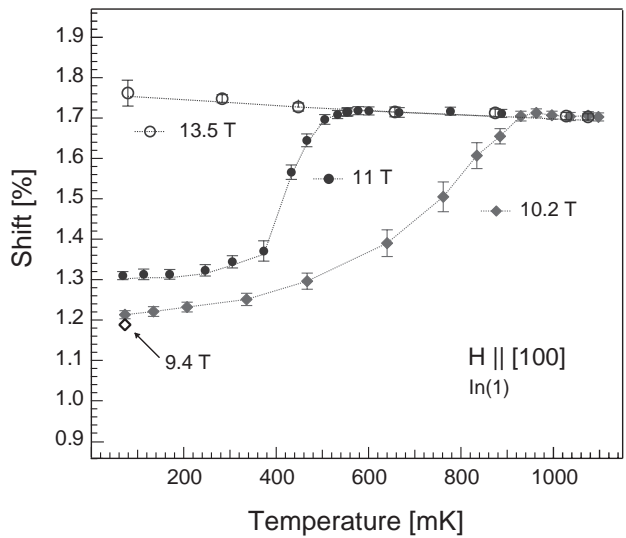}} 
 \vspace*{-0.2cm}
\caption[]{\label{Fig2}\small The magnetic shift of $^{115}$In(1)  as a function of  $T$ at  \mbox{10.2,  11, and 13.5  T}. The dotted lines are   guide to the eye.  
The open diamond is the shift at 9.4 T. }
\vspace*{-0.4cm}
\end{figure}

 To gain further insight into the nature of the FFLO state, 
  we proceed to the analysis of the spin shift.
In \mbox{Fig.  \ref{Fig2}} the $T$ dependence of the shift at 10.2, 11, and \mbox{13.5  T} is shown. 
In the high field normal state there is no evidence of any phase transition.
While  the 11 T data exhibits two phase transitions, our 10.2 T data reveals  that  the sample undergoes one phase transition to the uniform SC state. 
 In the uniform SC state at both 10.2 and \mbox{11 T}  the  decreasing shift with decreasing $T$
 reveals a well known suppression of the spin susceptibility consistent with spin-singlet pairing \cite{Curro01,Kohori}. Below $T^{\ast}$ at 11 T, the  $T$
  dependence 
 of the shift is strongly reduced. 
 Overall the spin shift seems to decrease  faster on approaching $T^{\ast}$ as compared to the its $T$  dependence in the uniform SC state. However,  this difference is not very significant on the reduced $T$ scale, as illustrated in the following paragraph.

 \begin{figure}[t]
 \centerline{\includegraphics[scale=0.47]{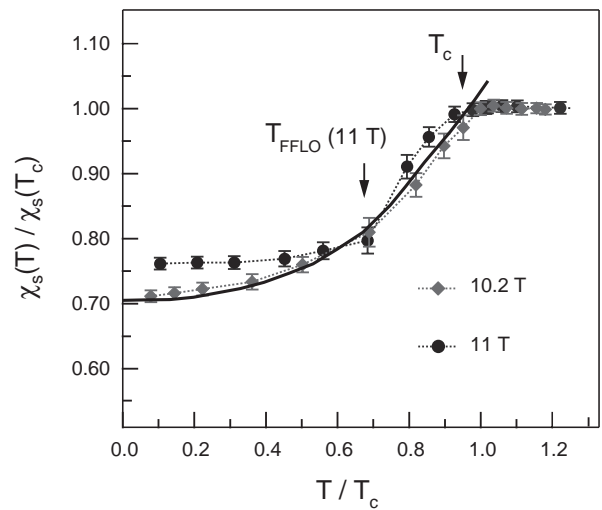}} 
 \vspace*{-0.2cm}
\caption[]{\label{Fig3}\small 
Normalized $^{115}$In(1) shift  in the SC states   as a function of $T$ at \mbox{10.2 and 11 T}.  
 Solid line denotes the quadratic $T$ dependence.   Transition temperatures
  (\Tc and $T_{\rm FFLO}  \equiv  T^{\ast}$) are adopted from Ref. \cite{Bianchi03} for comparison.}
  \vspace*{-0.5cm}
  \end{figure}

In   \mbox{Fig.  \ref{Fig3}},   the temperature dependence of the normalized shift (spin susceptibility) in the SC  states is plotted. 
 In the uniform SC state the  temperature  dependence of the local magnetization, which is proportional to the shift,  is expected to be quadratic, when $\mu_B B > k_BT$,  with the residual $T=0$ value due to the field induced shift in the spin-split density of states   \cite{AntonLT, KunYang}. 
 The solid line in  \mbox{Fig.  \ref{Fig3}} indicates such a $T$ dependence.  
  The data in the uniform SC state at both fields is in  agreement with the quadratic $T$ dependence. However,  below  \Tf 
 there is an obvious discrepancy, well outside the error bars, between data at 11 T  and the expected 
$T$ dependence in the uniform SC state.
The discrepancy is particularly pronounced at very low temperatures, {\it i.e.} $T \rightarrow 0$ limit.

 Therefore, we conclude that 
 in the state below 
 \Tf   {\it the spin shift is  enhanced} compared to  the  value in a uniform SC state, but reduced compared to the normal state. 
The enhancement 
is not a trivial effect due to increase of the applied field by 0.8 T on $\chi$, but it is a consequence of intrinsically different nature of the FFLO phase as compared to the uniform SC phase. 
This is because   in a uniform SC state  at 70 mK decreasing  $H$ from \mbox{10.2 T} by the same amount of \mbox{0.8 T} changes the  shift  by only \mbox{$\simeq 0.024$\%}.
This is 4 times less than the observed enhancement between 10.2 T and 11 T, as shown in   \mbox{Fig.  \ref{Fig2}}.

 Below $T^{\ast}$, we do not observe any  signal   at frequencies corresponding to the normal state signal.    
Therefore,  a  low $T$ state in our sample 
cannot simply be viewed as a stack of spatially well separated   SC  and normal regions, static on the NMR time scale. 
  However, this does not rule out the FFLO nature of the low  $T$  phase.  
  The spin shift can be enhanced by a  more  complex periodic superposition of 
 SC and ``normal'' regions, defined by a spectrum of spin-polarized quasiparticles
  \cite{Maki02}. Indeed,  calculations  indicate that a finite quasiparticle density of states (DOS)   is induced   at the Fermi level in the FFLO state \cite{Maki02,AntonPRB}. This finite DOS enhances the spin susceptibility, since $\chi \propto DOS(E_F)$. Furthermore, 
  the $\chi$ is found to be $T$ independent in the FFLO phase for $T$  below 
  \mbox{$\sim {1 \over 2} T^{*}$}  \cite{AntonLT, AntonPRB},  in agreement with 
our data.  As a matter of fact,
 our measurements are consistent with two possible pictures of the FFLO state: static and dynamic.
In the former, the FFLO state can be viewed as a static periodic modulation of the SC and ``normal'' regions, in which case 
the data imposes an important constraint on the local magnetization of the   ``normal'' quasiparticle regions. Mainly, it must be {\it significantly lower than the magnetization of the normal state}, since  no signal   is observed at frequencies corresponding to the normal state signal. 
 Alternatively in a dynamic scenario, the modulation of the amplitude of the order parameter in the FFLO state can fluctuate in space over a distance of the order of $1/q$. Thus, a measured shift would represent a volume weighted average  of the shifts in the SC and ``normal'' 
  regions.  In such a state, the NMR shift is higher than in the uniform SC state and no distinct signal from the normal regions should be observed, consistent with our  measurements.

    The observed  enhancement of the spin shift  could be induced by the appearance of  a small  static component   (ferromagnetic or canted antiferromagnetic)   in the  spin degrees of freedom.  
    However, there is no evidence from our high field data (at 13.5 T in \mbox{Fig. \ref{Fig2}}) that a 
    magnetic phase transition connected with an extension of the FFLO phase takes place  in the  normal state. Therefore, the existence of the magnetic moment requires    pair coherence.     
    In a more exotic scenario, 
	the  low $T$ phase could be a  mixture of singlet and triplet pairs. It is possible that in 
	high magnetic field  it becomes energetically favorable to form triplets through the ferromagnetic  coupling channel \cite{CooperPrivate}, even though the states of mixed parity are not {\it a priori}  allowed in a material with inversion center. 
Finally,   $\chi$ may also be enhanced  by   the uncondensed electrons that  coexist with the SC ones on a microscopic lengthscale, so that no separate signal is observed at the frequencies corresponding to the normal state signal. The existence of uncondensed electrons could be explained within  the context of multiband SC  \cite{tanatar05}.
 
 There are evident discrepancies between our data and the results reported  in
  \mbox{Ref. \cite{Kakuyanagi05}}. Besides the lack of the  signal at frequencies corresponding to the normal state signal, the relative shift changes  are inconsistent.   The normal state $^{115}$In(1) shift   \cite{shiftCom} presented here  is comparable to the shifts reported  in  \mbox{Ref. \cite{Curro01, Kakuyanagi05}}.  
However, the  relative shift changes between the low temperature SC and  normal state, evident  in \mbox{Fig.  \ref{Fig2}},  are $\simeq 10$ times larger than those reported in   \mbox{Ref. \cite{Kakuyanagi05}}.  Furthermore, the transition temperatures found in this work are in agreement with specific heat measurements of  \mbox{Bianchi {\it et al.}}   \cite{Bianchi03} and different from those observed in \mbox{Ref. \cite{Kakuyanagi05}}.  It is unlikely that the sample quality is the origin of these discrepancies, since the spectral lineshapes and their corresponding widths in high $T$ normal state and at the lowest $T$ are comparable. 
It is possible that differences stem from  the exact sample alignment with respect to $H$  
\cite{Radovan03}.

In conclusion, our NMR study of  In(1) magnetic shift in \Ce   in a magnetic field applied parallel to SC  planes  has confirmed the existence of two phase transitions at low $T$ in the vicinity of $H_{\rm c2}$. 
At 11 T, we show that   the   phase transition at \Tc     clearly corresponds to  a  transition to the uniform  spin-singlet SC state. 
We find that the main feature of the   low $T$ state below \Tf is the    enhancement of   the spin susceptibility as compared to  the  value in a uniform SC state.  In the FFLO scenario for the low $T$ state, our data  supports two possible pictures. First, the static FFLO state, in which the local magnetization of the   ``normal''  regions  must be significantly lower than the magnetization of the normal state. Second, the dynamic FFLO state in which the modulation of the amplitude of the order parameter  fluctuates in space. 
 Alternatively, it is possible that  the low $T$ phase arises from an intricate interplay of magnetism and superconductivity.

We are very grateful to V. Mineev,  {\v C.} Petrovi{\' c},  A. Benlagra, A. B. Vorontsov, M. J. Graf,  
 J. P. Brison,  S. Kr\"{a}mer, and A. J. Sauls for helpful discussions.  Special thanks to W. P. Halperin for the magnet time and useful discussions. 
This work was supported by the funds from  Brown University, and the Grenoble High Magnetic Field Laboratory, under European Community contract RITA-CT-2003-505474. 
 
\vspace{-0.4cm}
\bibliographystyle{unsrt}

\begin{thebibliography}{10}
\bibitem{shiftCom}
There is a discrepancy of $\simeq 0.3\%$  between the normal state shift at   low $T$ reported here and in  \mbox{Ref. \cite{Kakuyanagi05}}. It is in part due to our subtraction of the orbital contribution to the  shift; and, in part to  the lack of  knowledge (leading to significant error in shift \cite{Curro01}) of the exact  tilt angle between $H$ and the samples  $(\hat a \hat b)$ planes in \mbox{Ref. \cite{Kakuyanagi05}}.
\end{thebibliography}

\end{document}